# Unidirectional scattering with spatial homogeneity using photonic time disorder

Jungmin Kim[1,2], Dayeong Lee[2], Sunkyu Yu[2,†], and Namkyoo Park[1,*]

[1]Photonic Systems Laboratory, Department of Electrical and Computer Engineering, Seoul National University, Seoul 08826, Korea

[2]Intelligent Wave Systems Laboratory, Department of Electrical and Computer Engineering, Seoul National University, Seoul 08826, Korea

E-mail address for correspondence: †sunkyu.yu@snu.ac.kr, *nkpark@snu.ac.kr

**Abstract**

The temporal degree of freedom in photonics has been a recent research hotspot due to its analogy with spatial axes, causality, and open-system characteristics. In particular, the temporal analogues of photonic crystals have stimulated the design of momentum gaps and their extension to topological and non-Hermitian photonics. Although recent studies have also revealed the effect of broken discrete time-translational symmetry in view of the temporal analogy of spatial Anderson localization, the broad intermediate regime between time order and time uncorrelated disorder has not been examined. Here we investigate the inverse design of photonic time disorder to achieve optical functionalities in spatially homogeneous platforms. By developing the structure factor and order metric using causal Green's functions for the domain of time disorder, we demonstrate engineered time scatterer, which provides unidirectional scattering with controlled scattering amplitudes. We also reveal that the order-to-disorder transition in the time domain allows for the manipulation of scattering bandwidths, which inspires resonance-free temporal colour filtering. Our work will pave the way for advancing optical functionalities without spatial patterning.

## Introduction

Associating temporal and spatial axes has enriched the perspective on manipulating wave phenomena. Owing to the space-time analogy between the electromagnetic paraxial equation and the Schrödinger equation, the temporal axis can be considered an alternative or auxiliary axis to the spatial dimension. This similarity between temporal and spatial axes has established the fields of quantum-optical analogy[1], non-Hermitian[2], topological[3,4], and supersymmetric[5,6] photonics, and universal linear optics[7]. On the other hand, the uniqueness of a temporal axis has also been a recent research focus for achieving distinct design freedom from spatial freedom[8,9]. For example, the broken time-translational symmetry results in dynamical wave responses, composing open systems. In this context, dynamical wave devices with optical nonlinearity[10,11] or non-Markovian processes[12] require the design strategy to appropriately break the time-translational symmetry. Furthermore, causality leads to unique scattering distinct from its spatial counterpart, completely blocking backscattering along the temporal axis[13].

Recent studies utilizing temporal degrees of freedom have thus focused on exploiting similarities and differences between temporal and spatial axes. The discrete time-translational symmetry in photonic time crystals (PTCs)[14,15] has been examined as a temporal analogy of photonic crystals, revealing the unique phenomena along the temporal axis, such as momentum bandgaps and the localized temporal peak due to the Zak phase. The concept of disordered photonics has also been extended to the temporal axis, such as observing the statistical amplification and the scaling of Anderson localization in uncorrelated disorder[16,17]. Various wave physics, such as amplification and lasing[18,19], effective medium theory[20], Snell's law[13], spectral funnelling[21], supersymmetry[22], parity-time symmetry[23], nonreciprocity[24], and metamaterials[25-28], have also revealed the unique features and applications of the temporal axis inspired by its spatial counterparts. Nonetheless, these intriguing achievements cover only the partial regimes in microstructural statistics of temporal modulations, such as order with conserved symmetries[14,15,18,19,22,23,29], and their breaking with finite defects[13,24] or perturbations without any correlations[16,17]. When considering abundant degrees of freedom in material microstructures[30], further attention to the intermediate regime between order and uncorrelated disorder for the temporal axis is mandatory.

In this paper, we propose the concept of engineered time disorder, which allows for the designed manipulation of light scattering. Starting from the theoretical framework for analysing spatial disorder, we build its temporal analogue by incorporating causality in the time axis, which allows for examining the relationship between the time structure factor, time-translational order metric, and wave scattering. We demonstrate that the moulding of the structure factor enables the completely independent engineering of forward and backward scattering. By investigating the order-to-disorder transition in the temporal modulation of the system, we also enable bandwidth engineering of unidirectional scattering, such as time disorder for broadband scattering and resonance-free colour filtering. Our result verifies the spatial-pattern-free design of conventional optical functionalities and represents a great advantage of time disorder in bandwidth engineering with respect to time crystals.

## Results

**Temporal scattering**. Consider a nonmagnetic, isotropic, and spatially homogeneous optical material having the time-modulated relative permittivity $\varepsilon(t)$. For the $x$-polarized planewave of the displacement field $\mathbf{D}(\mathbf{r},t) = \mathbf{e}_x\psi(t)e^{ikz}$, where $k$ is the wavenumber, the governing equation is[6,14,16]:

$$\left[\frac{d^2}{dt^2} + \frac{c^2 k^2}{\varepsilon(t)}\right]\psi(t) = 0, \tag{1}$$

where $c$ is the speed of light. Because $k$ is conserved according to the spatial translational symmetry, Eq. (1) is the temporal analogy of the one-dimensional (1D) Helmholtz equation for spatially varying materials, exhibiting space-time duality[28] by imposing the role of the optical potential on $[\varepsilon(t)]^{-1}$. To investigate the regime of weak scattering, we express the real-valued optical potential as $\alpha(t) \equiv \varepsilon^{-1}(t) = \alpha_b[1 + \Delta\alpha(t)]$, where $\alpha_b$ is the potential at $t \to \pm\infty$. With the assumption of weak perturbation during the finite temporal range, the time-varying component $\alpha_b \Delta\alpha(t)$ becomes analogous to the weakly perturbed permittivity in spatial-domain problems[1]. Notably, as those time-varying systems are open systems, the energy provided by the environment $P_{in}(t) = du_{EM}^0/dt$ results in the nonconservative EM field energy[14,16] $u_{EM}^0(t) = [\mathbf{E}^*(t) \cdot \mathbf{D}(t) + \mathbf{H}^*(t) \cdot \mathbf{B}(t)]/4$.

For a given temporal variation of the system, we employ the harmonic incidence $\psi_{inc}(t) = \exp(-i\omega_b t)$, where $\omega_b = \alpha_b^{1/2} kc$ is the optical frequency at $t \to \pm\infty$. Under the first-order Born approximation[31] with $|\Delta\alpha(t)| \ll 1$, the time-domain scattering field $\psi_{sca}(t)$ becomes:

$$\psi_{sca} \simeq -\omega_b^2 \int_{-\infty}^{\infty} dt' \Delta\alpha(t') \psi_{inc}(t') G(t; t'), \tag{2}$$

where $G(t; t')$ is the Green's function for the impulse response of the temporal delta function scatterer $\delta(t - t')$ (Supplementary Note S1).

Although Eq. (2) is identical to the 1D scattering problem in the spatial domain[31], the uniqueness of the time-varying material is in the mathematical form of the Green's function: the concept of the Feynman and retarded propagators[32]. Due to the unidirectional flow of time, the temporal Green's function satisfies causality: $G(t; t') = 0$ for $t < t'$. To fulfil the temporal boundary conditions for the displacement field and magnetic field[9,14] at $t = t'$, the analytical form of the retarded Green's function for the temporal impulse becomes (Fig. 1a, Supplementary Note S2):

$$G(t;t') = \frac{1}{\omega_b} \sin[\omega_b(t-t')]\Theta(t-t'), \tag{3}$$

where $\Theta(t)$ is the Heaviside step function of $\Theta(t > 0) = 1$ and $\Theta(t \leq 0) = 0$ (Supplementary Note S3). The Green's function in Eq. (3) can be separated into $G(t; t') = G_{FW}(t; t') + G_{BW}(t; t')$ for $G_{FW,BW}(t; t') = \pm\exp[\pm i\omega_b(t-t')]\Theta(t-t')/(2i\omega_b)$, where each sign of $\pm\omega_b$ determines the propagation direction with the conserved $k$.

We emphasize that causality imposes the uniqueness on the temporal Green's function, i.e., the coexistence of the forward (or transmitted) and backward (or reflected) waves in $t > t'$ (Fig. 1a). Such a mathematical form of $G(t; t')$ is in sharp contrast to the spatial Green's function $G(z; z') \sim \exp(ik|z - z'|)$, which exhibits the separate existence of the forward ($e^{+ik(z-z')}$) and backward ($e^{-ik(z-z')}$) waves in $z > z'$ and $z \leq z'$, respectively (Fig. 1b). This uniqueness emphasizes the open-system nature of time-varying systems, despite the fact that the governing equation of Eq. (1) is mathematically identical to the spatial one. When an external modulation to the system is applied by a time-varying signal power $P_{in}$ (Fig. 1c), the unique form of the causal Green's function in a time-varying system—interfering forward and backward basis (Fig. 1a)—breaks the conservation of the electromagnetic energy inside the system (Fig. 1d). In this context, the independent control of forward and backward scattering in temporally random heterogeneous materials compels a design strategy distinct from their spatial counterparts.

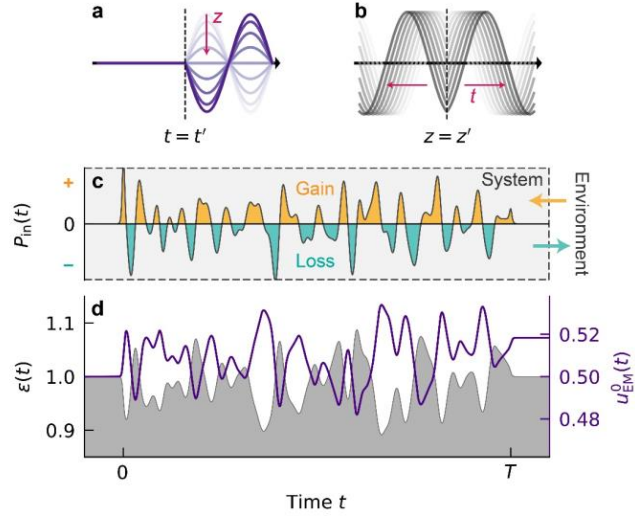

**Fig. 1. Concept of temporal scattering as open-system responses. a, b,** Schematics of temporal and spatial Green's functions: (**a**) Re[$G(t; t')e^{+ikz}$] and (**b**) Re[$G(z; z')e^{-i\omega t}$]. Shaded lines in (**a, b**) indicate the evolution of each Green's function. **c,** Schematic of system modulation by signal power $P_{in}(t)$ from the environment, representing the system gain and loss for positive and negative $P_{in}$, respectively. **d,** Energy alteration from light-matter interactions with the time disorder driven by $P_{in}(t)$ in (**c**). $\varepsilon(t)$ (grey area) and $u_{EM}^0(t)$ (purple line) are the time-varying permittivity confined inside the temporal range [0, $T$] and the instantaneous electromagnetic energy density, respectively.

From Eqs (2) and (3), the scattering field becomes (Supplementary Note S4):

$$\psi_{sca}(t) = \frac{\omega_b}{2i}\left[e^{-i\omega_b t}\int_{-\infty}^{t}dt'\Delta\alpha(t') - e^{+i\omega_b t}\int_{-\infty}^{t}dt'e^{-2i\omega_b t'}\Delta\alpha(t')\right]. \quad (4)$$

In Eq. (4), we assume a finite-range temporal variation $\Delta\alpha(t)$ with $\Delta\alpha(t < 0) = \Delta\alpha(t > T) = 0$. We also employ the ergodic hypothesis, i.e., the statistical equivalence between the average over all realizations and the average over one statistically homogeneous realization at the thermodynamic limit[33]. The ergodicity allows for the homogeneous correlation function $\hat{S}(t_1, t_2) \equiv \langle\Delta\alpha^*(t_1)\Delta\alpha(t_2)\rangle = S(\Delta t)$ for $0 \leq t_{1,2} \leq T$ and $\Delta t = t_1 - t_2$, where $\langle.\rangle$ denotes the ensemble average. We separate the ensemble-averaged scattering power after the temporal perturbation ($t > T$) into the forward ($\langle P_{FW}\rangle$) and backward ($\langle P_{BW}\rangle$) waves (Supplementary Note S5) as:

$$\begin{aligned}\langle P_{\text{FW}}\rangle &= \frac{\omega_b^2}{4}\int_0^T dt_1'\int_0^T dt_2' S(t_1'-t_2'),\\ \langle P_{\text{BW}}\rangle &= \frac{\omega_b^2}{4}\int_0^T dt_1'\int_0^T dt_2' S(t_1'-t_2')e^{2i\omega_b(t_1'-t_2')}.\end{aligned} \quad (5)$$

With a sufficiently broad temporal range, each power approaches the Fourier transform of $S(\Delta t)$ (Supplementary Note S6), as follows:

$$\langle P_{\text{FW}}\rangle \approx \frac{\omega_b^2 T}{4}S(0), \quad \langle P_{\text{BW}}\rangle \approx \frac{\omega_b^2 T}{4}S(2\omega_b), \quad (6)$$

where we define $S(\omega)$ as the "time structure factor" governing scattering from temporal disorder, *i.e.*, the temporal counterpart of the static structure factor[33,34]. Therefore, engineering the power flows $P_{\text{FW,BW}}$ using time disorder is achieved by moulding $S(\omega)$ near $\omega = 0$ and $2\omega_b$. Notably, in the design of $S(\omega)$, three conditions should hold for $S(\Delta t)$ and $S(\omega)$: the Hermiticity $S(\Delta t) = S^*(-\Delta t)$ with real-valued $S(\omega)$, $|\text{Re}[S(\Delta t)]| \leq S(\Delta t = 0)$ from the maximum of the correlation function, and $S(\omega) \geq 0$ from the autocorrelation theorem (Supplementary Note S7).

To establish the designed manipulation of light through photonic time disorder, we demonstrate the engineering of time disorder: unidirectional scattering for the independent control of $\langle P_{\text{FW}}\rangle$ and $\langle P_{\text{BW}}\rangle$, order-to-disorder transition for spectral manipulation, and momentum-selective spectral shaping. We note that there are two different classes of one-to-many correspondence between a scattering response and the realizations of disorder. First, because scattering phenomena are governed by $S(0)$ and $S(2\omega_b)$ for a planewave of $\omega_b$, a family of time disorder can be achieved by altering the overall shape of $S(\omega)$ while preserving $S(0)$ and $S(2\omega_b)$. Second, even for the same $S(\omega)$, there is an infinite number of realizations of time disorder because the entire landscape is uniquely determined by all the orders of correlation functions[33]. We study both origins in examining a family of disorder in the following discussion.

**Unidirectional scattering.** We demonstrate the unidirectional scattering achieved by suppressing $S(0)$ or $S(2\omega_b)$. We set a time scale $t_0$ and assume the incident frequency $\omega_b = \omega_0$. We set the structure factor functions $S_{FW}(\omega)$ and $S_{BW}(\omega)$ for the forward and backward scatterings, which are designed in the frequency ranges $[-2\omega_0, 2\omega_0]$ and $[-3\omega_0, 3\omega_0]$, respectively, and zero elsewhere. The structure factors $S_{FW,BW}(\omega)$ in this scenario are modulated by the design parameters $S_0 = S_{FW}(0)$ and $S_{2\omega} = S_{BW}(2\omega_b)$, respectively, while satisfying the continuity and $C^1$ smoothness as well as the statistical bounds for $\varepsilon(t)$ (Supplementary Note S8).

Figure 2a shows the designed $S_{FW}(\omega)$ and $S_{BW}(\omega)$ for different values of $S_0$ and $S_{2\omega}$, respectively. With the corresponding $S(\Delta t)$ from $S_{FW,BW}(\omega)$, we generate a set of $\varepsilon(t)$ realizations through the multivariate Gaussian process (Supplementary Notes S7 and S9). Three example realizations are depicted in Fig. 2b for the suppressions of (i) both forward and backward (case A), (ii) backward only (case B), and (iii) forward only (case D), which have the corresponding time structure factor shown in Fig. 2a.

For each case, an ensemble of $10^4$ realizations is generated, and their scattering responses are examined using the time-domain transfer matrix method (TD-TMM)[16,35]. Figures 2c, d show that the ensemble average of the rigorous TMM results (error bars) provides good agreement with the $S(\omega)$-based prediction with the Born approximation (lines). Engineering temporal modulation using the time structure factor allows for completely independent manipulation of temporal scattering: unidirectional scattering only with forward (case B) or backward (case D) propagations or scattering-free temporal variations (cases A and C).

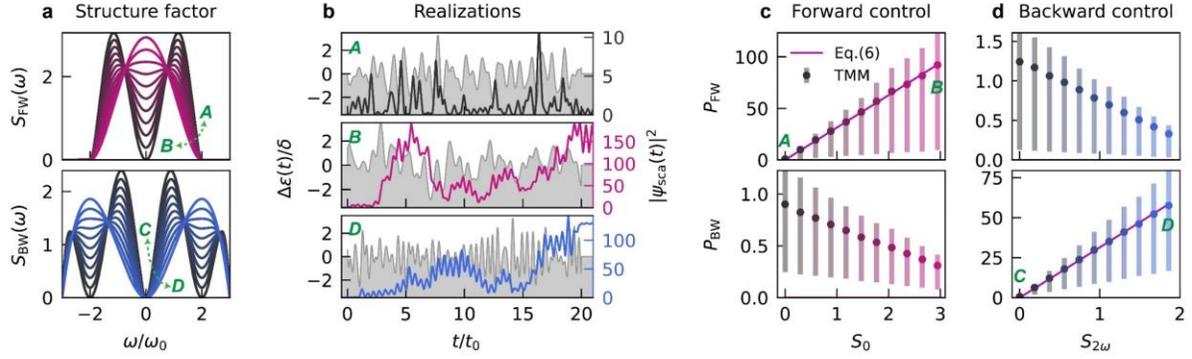

**Fig. 2. Engineering unidirectional scattering. a,** Structure factors $S_{FW}(\omega)$ (top) and $S_{BW}(\omega)$ (bottom) with varying design parameters $S_0$ and $S_{2\omega}$, respectively. The design parameters are represented by gradual colours (A: $S_0 = 0$, B: $S_0 \sim 2.95$, C: $S_{2\omega} = 0$, and D: $S_{2\omega} \sim 1.86$). **b,** An example of the $\Delta\varepsilon(t)$ realizations (grey areas) and the corresponding scattering intensity $|\psi_{sca}(t)|^2$ (solid lines) for the A, B, and D states in (**a**). $t_0 = 2\pi/\omega_0$. **c, d,** Comparison of the scattering powers from the structure factor prediction (solid lines) and rigorous TD-TMM (error bars) for each ensemble ($10^4$ realizations) with different design parameters $S_0$ and $S_{2\omega}$ in (**a, b**): the suppression of the (**c**) backward and (**d**) forward scattering. The top and bottom figures represent the forward and backward power, respectively, after $t > 20t_0$. Markers denote the ensemble average from the TD-TMM results. Error bars denote the 1$^{st}$ to 3$^{rd}$ quartiles of the ensemble. Structure factors [$S_{FW,BW}(\omega)$, $S_{0,2\omega}$], scattering field ($\psi_{sca}$), and scattering powers ($P_{FW,BW}$) are normalized with $\delta^2/\omega_0$, $\delta$, and $\delta^2$, respectively, where $\delta = [S(\Delta t = 0)]^{1/2}$.

**Engineered time disorder for spectral manipulation.** The main advantage of utilizing disordered systems in wave physics is the ability to manipulate multiple wave quantities with different sensitivities to material phases[30]. Such intricate wave-matter interactions allow for the alteration of the target wave quantity while preserving other ones, as shown in the independent manipulation of localization and spectral responses in spatial domains[6,36]. In this context, we focus on the independent control of two wave properties—scattering and spectral responses—using photonic time disorder.

In designing temporal systems through the language of the time structure factor, ordered systems (*e.g.*, photonic time crystals[15]) are depicted by a set of Bragg peaks, indicating certain harmonic frequencies at which the system interacts with an incident wave. In contrast, time disorder close to the Poisson process[37] shows a broadband structure factor that guarantees a continuum frequency response; at the extreme, the uncorrelated Poisson disorder possesses the structure factor of an infinite plateau. Using such a clear distinction between order and uncorrelated disorder and the relationship between the structure factor and scattering, we explore the intermediate regime between two extremes in photonic time disorder.

To quantify the transition between order and uncorrelated disorder, we introduce the transition parameter $\xi$ for the structure factor $S(\omega)$: from $\xi = 0$ mimicking crystals to $\xi = 1$ for a near-Poisson case. We set the extreme case of the structure factors $S_C(\omega)$ and $S_P(\omega)$ for the crystal and near-Poisson state, respectively, defining the transition between them, as (Fig. 3a, Supplementary Note S10)

$$S(\omega) = (1-\xi)S_C(\omega) + \frac{1}{2}\left[S_P\left(\frac{\omega - 2\omega_0}{\xi}\right) + S_P\left(\frac{\omega + 2\omega_0}{\xi}\right)\right]. \tag{7}$$

Figure 3b shows the structure factors obtained from different mixing of $S_C(\omega)$ and $S_P(\omega)$, targeting the suppression of forward power $P_{FW} \sim 0$ with $S(\omega = 0) = 0$. As the transition from the A to D states occurs, the heights of the Bragg peaks from $S_C(\omega)$ at $\omega \neq 2\omega_0$ decrease, while the bumps $S_P(\omega)$ centred at $\omega = \pm 2\omega_0$ (Fig. 3b, inset) are continuously broadened. Equation (7) allows for maintaining the integral of $S(\omega)$ over the frequency domain to restrict the average fluctuation in the time domain realizations. The designed transition therefore enables the characterization of time disorder solely depending on the "pattern" of disorder, not on the magnitude of the fluctuation.

Figure 3c shows examples of the realization of time disorder for different $\xi$ values in Fig. 3b, all of which are designed to derive backward scattering only. The transition parameter $\xi$ qualitatively describes the temporal material phase transition from nearly crystalline to nearly uncorrelated disorder. To characterize each disorder more quantitatively, we introduce the time-translational order metric $\tau$:

$$\tau = t_0^{-1} \int_{-4\omega_0}^{4\omega_0} d\omega \left| S(\omega) - \frac{\pi \delta^2}{4\omega_0} \right|^2, \tag{8}$$

where $\pm 4\omega_0$ denotes the range of nonzero $S(\omega)$ (Supplementary Note S10). Analogous to its original definition in the spatial domain[34,37], $\tau$ characterizes the distance of a given temporal evolution from the Poisson process, describing how much a given $\varepsilon(t)$ is ordered in the time domain. As shown in Fig. 3d, the designed backward scattering from Eq. (6) and the resulting TD-TMM show good agreement, while better agreement is achieved when we leave out the infinite temporal range approximation by using Eq. (5).

The most significant difference between crystals and uncorrelated disorder can be found in their spectral responses. As shown in Figs. 3e, f, the change in material phases between order and uncorrelated disorder provides a designed manipulation of the bandwidth of temporal modulations while preserving the target scattering response: the suppression of forward scattering. Remarkably, near-Poisson time disorder (case D) guarantees almost ±10% range of spectrum bandwidth for the suppression of forward scattering and the constant backward scattering, improving the bandwidth 40 times compared to that of the near-crystal one (case A). Therefore, the use of randomness in the temporal modulation enables a significant bandwidth enhancement and thus the noise-robust signal processing preserving optical functionalities.

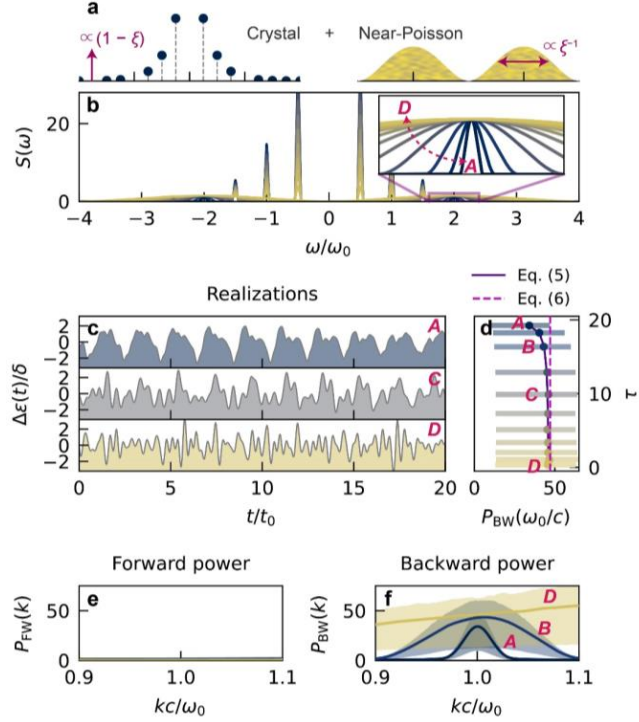

**Fig. 3. Time disorder for bandwidth engineering**. **a,** Schematics of the structure factors for crystalline (left) and near-Poisson disorder (right). $\xi$ is the transition parameter between order ($\xi = 0$) and disorder ($\xi = 1$). **b,** Structure factors $S(\omega)$ for the order-to-disorder transition with varying $\xi$: from crystalline (navy) to near Poisson (yellow). A, $\xi = 0.025$; B, $\xi = 0.1$; C, $\xi = 0.3$; D, $\xi = 1$. **c,** Examples of realizations of $\Delta\varepsilon(t)$ for A, C, and D. **d,** Statistical relationship between the backward scattering power and the time-translational order metric $\tau$. The scattering theory prediction (solid and dashed lines with Eqs (5) and (6), respectively) and rigorous TD-TMM (error bars) are compared for each ensemble of $10^4$ realizations. **e, f,** Spectral responses of the (**e**) forward and (**f**) backward scattering powers near the target momentum ($0.9 < kc/\omega_0 < 1.1$) for A, B, and D in (**d**). Solid lines and coloured areas denote the ensemble averages and the 1st to 3rd quartiles, respectively. Structure factors [$S(\omega)$], order metric ($\tau$), and scattering powers ($P_{FW,BW}$) are normalized with $\delta^2/\omega_0$, $\delta^4$, and $\delta^2$, respectively, where $\delta = [S(\Delta t = 0)]^{1/2}$.

**Momentum-selective spectral shaping.** Through Figures 2 and 3, we demonstrate the control of the scattering directivity under monochromatic conditions and its spectral engineering through an order-to-disorder transition. Based on this result, we show a novel methodology for the filtering of light waves—temporal "resonant-less" colour filter—using a platform with spatial translational symmetry. The proposed approach is in sharp contrast to conventional platforms for light filtering, such as multilayers[38] or resonators[39].

As an example of this application, we consider the propagation of a pulse and its interaction with the designed photonic time disorder, which leads to unidirectional and bandpass scattering. Because the forward scattering is governed by $S(0)$ regardless of the light momentum $k$, the momentum-resolved operation for the forward scattering is prohibited. Therefore, we focus on the filtering of backward waves while suppressing forward waves, as illustrated in Fig. 4a, which filters out the range of "wave" momenta $k$ by the corresponding "material" temporal frequencies $|\omega| = 2c|k| \in [\omega_0/2, \omega_0]$. Notably, we set $S(\omega) \sim \omega^{-2}$ dependency in the target range to compensate for the $\omega^2$ dependency of scattering power [Eq. (6)]. The temporal correlation and a sample realization of a given structure factor are shown in Figs. 4b, c.

The initial displacement field $D(z, t = 0)$ is a real-valued scalar function that satisfies $D(k) = D^*(-k)$. We assume a Gaussian pulse $D(z, t = 0) = \exp[-(z/\sigma_z)^2/2]$. The time evolution of the field through the time disorder filter becomes

$$D(z,t) = \int_{-\infty}^{\infty} \frac{dk}{2\pi} D(k, t=0) e^{ikz} \psi_{\text{tot}}(t; k), \qquad (9)$$

where $\psi_{\text{tot}}(t; k)$ is the single-component response of the incident planewave $\psi_{\text{inc}}(t) = e^{-ikct}$. Snapshots of the pulse evolution are shown in Fig. 4d, exhibiting $+z$ propagation (pink arrow) and a scattered tail behind it.

The evolutions of scattered fields in real- and $k$-space are illustrated in Figs 4e and f, respectively. After the modulation, the generated backscattered field propagates along the $-z$-direction (blue arrow). Figure 4f and its $\omega$-axis representation (Fig. 4g) clearly demonstrate the filtering functionality, which preserves the envelope shape of the original incident pulse while suppressing the designed band stop range $ck/\omega_0 \in [-1/4, 1/4]$ (Supplementary Movie S1).

The mechanism of the suggested temporal colour filter is fundamentally distinct from conventional optical filters, which utilize the bounded momentum responses through spatial inhomogeneity (for example, using mirrors, scatterers, or resonators) and the following constraint on spectral responses through dispersion relations. In contrast, the proposed temporal colour filter does not require spatial inhomogeneity. While the momentum of light is preserved through spatial translational symmetry, the spectral responses are filtered through broken temporal translational symmetry, which is the nature of time-varying open systems.

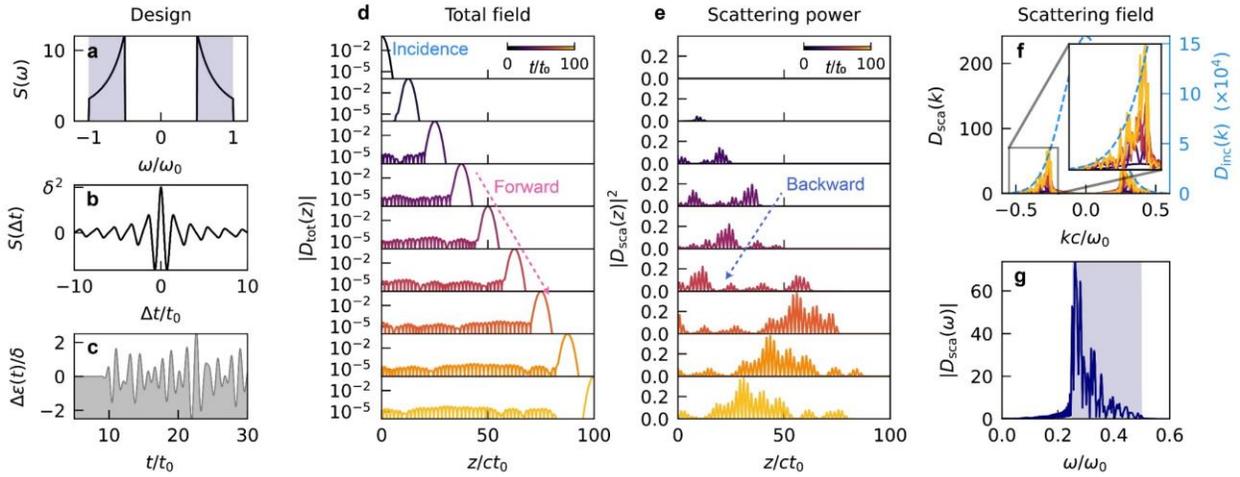

**Fig. 4. Momentum-selective scatterer**. **a,** Target structure factor for momentum-dependent spectral shaping of backscattering. **b**, **c,** The corresponding (**b**) temporal correlation function and (**c**) sample realization of $\Delta\varepsilon(t)$. **d**, **e,** Time evolution of a Gaussian pulse $D_{inc}(z, t = 0) = \exp[-(z/\sigma_z)^2/2]$ undergoing a tailored temporal perturbation from $t/t_0 = 10$ to $90$: (**d**) the total field amplitude and (**e**) scattering power. Arrows indicate the direction of propagation for the incident and back-reflected fields. **f**, **g,** Spectral responses of the scattering field: (**f**) the time evolution of the $k$-space field $D_{sca}(k, t)$ and (**g**) the $\omega$-domain field $D_{sca}(z = 0, \omega)$ at a fixed point. The blue dashed line in (**f**) denotes $D_{inc}(z, t = 0)$. The shaded area in (**g**) represents the filtering band $S(2ck)$. The gradual colours in (**d**-**f**) represent the time evolutions from $t = 0$ (black) to $100t_0$ (yellow). Structure factors $S(\omega)$ and fields $D_{sca}(z)$, $D_{sca,inc}(k)$, and $D_{sca}(\omega)$ are normalized with $\delta^2/\omega_0$, $\delta$, $\delta c/\omega_0$, and $\delta/\omega_0$, respectively, where $\delta = [S(\Delta t = 0)]^{1/2}$.

**Discussion**

Recently, several important studies have explored time disorder[16,17], revealing the growth of the statistical intensity of waves with log-normal distributions and temporal Anderson localization, both of which are obtained with uncorrelated disorder. In contrast, the significance of our result is on bridging temporal light scattering and correlated time disorder, which allows for the deterministic engineering of scattering direction, bandwidth, and spectral shaping.

In terms of time-dependent perturbation theory, our statistical approach corresponds to a weak perturbation with the spectral transition amplitude $S(\omega)$ that gives rise to the transition from the incident (initial state, $\omega_b$) to forward or backward scattering (1$^{st}$-order perturbations, $\pm\omega_b$) with the energy differences $S(\Delta\omega = -2\omega_b, 0)$, respectively, as similar to Fermi's golden rule. Furthermore, this is consistent with a so-called time correlation function in the Green-Kubo (G-K) relationship[40,41] to describe the transport coefficient in fluid[42] or thermal[43] systems. While the listed phenomena all share the universal linear response theory in both classical and quantum physics, the direct application of the essence of the G-K relation to temporal light scattering is demonstrated for the first time, and it serves as a toolkit for dynamical photonic systems.

Notably, time-varying wave systems can be experimentally realized through time-varying transmission lines (TVTLs)[9,18,44-47]. Because transmission lines are ideal platforms for describing one-dimensional wave propagations, TVTLs with temporal modulations via loaded LC resonators or varactor diodes allow for reproducing intriguing phenomena in time-varying wave systems. For the practical implementation of our disordered systems, the TVTL is also expected to be a suitable platform, which only requires the free-form control of time-varying parameters.

To summarize, we developed the patternless realization of EM scattering in temporally disordered media. Starting from the analytical formulation of wave scattering with the time structure factor, we demonstrated the moulding of the structure factors for engineered scattering. This top-down approach enables the design of modulation signals for unidirectional scattering in spatially homogeneous systems. By examining the order-to-disorder transition in the temporal domain, we also develop bandwidth engineering while preserving unidirectional scattering, which enables the realization of resonance-free colour filters. In terms of engineered disorder, exploring nontrivial material states defined in the temporal domains, such as hyperuniformity[34] and stealthy[30], will be a further research topic.

**Methods**

**Data availability**

The data that support the plots and other findings of this study are available from the corresponding author upon request.

**Code availability**

All code developed in this work will be made available from the corresponding authors upon request.

**Acknowledgements**

This work was supported by the National Research Foundation of Korea (NRF) through the Basic Research Laboratory (No. 2021R1A4A3032027), Young Researcher Program (No. 2021R1C1C1005031), and Global Frontier Program (No. 2014M3A6B3063708), all funded by the Korean government.

## Author contributions

J.K., S.Y., and N.P. conceived the idea. J.K. developed the theoretical tool and performed the numerical analysis. J.K. and D.L. examined the theoretical and numerical analysis. All the authors discussed the results and wrote the final manuscript.

## Competing interests

The authors have no conflicts of interest to declare.

## Additional information

**Correspondence and requests for materials** should be addressed to S.Y. (sunkyu.yu@snu.ac.kr) or N.P. (nkpark@snu.ac.kr).

# Supplementary Information for

# "Unidirectional Scattering with Spatial Homogeneity using Photonic Time Disorder"

Jungmin Kim,[1,2] Dayeong Lee,[2] Sunkyu Yu,[2,†] and Namkyoo Park[1,*]

[1]Photonic Systems Laboratory, Dept. of Electrical and Computer Engineering, Seoul National University, Seoul 08826, Korea

[2]Intelligent Wave Systems Laboratory, Dept. of Electrical and Computer Engineering, Seoul National University, Seoul 08826, Korea

[†] *sunkyu.yu@snu.ac.kr*

[*] *nkpark@snu.ac.kr*

Note S1. Scattering with the Born approximation

Note S2. Causal Green's function

Note S3. Numerical validation of the Born approximation

Note S4. Directional separation of scattering waves

Note S5. Ensemble-averaged forward and backward powers

Note S6. Long-range approximation

Note S7. Gaussian random generation and the conditions for structure factors

Note S8. Design of structure factors for target forward and backward scatterings

Note S9. Estimation of $S(\omega)$ for generated realizations

Note S10. Details of $S_C(\omega)$ and $S_P(\omega)$

**Note S1. Scattering with the Born approximation**

With the perturbed potential $\alpha(t) \equiv \varepsilon^{-1}(t) = \alpha_b[1 + \Delta\alpha(t)]$, Eq. (1) in the main text is modified as follows:

$$\left(\frac{d^2}{dt^2} + \omega_b^2\right)\psi(t) = -\omega_b^2 \Delta\alpha(t)\psi(t). \tag{S1}$$

Assuming that the impulse response of the above operator $(d^2/dt^2 + \omega_b^2)$ is expressed as:

$$\left(\frac{d^2}{dt^2} + \omega_b^2\right)G(t,t') = \delta(t-t') \tag{S2}$$

with a Green's function $G(t, t')$, the total field $\psi_{tot}(t)$ as the solution of Eq. (S1) is given by the sum of the homogeneous solution $\psi_{inc}(t) = \exp(-i\omega_b t)$ (*i.e.*, the incident wave) and the inhomogeneous solution $\psi_{sca}(t)$ (*i.e.*, the scattered wave), where:

$$\left(\frac{d^2}{dt^2} + \omega_b^2\right)\psi(t) = -\omega_b^2 \Delta\alpha(t)\psi(t) = -\omega_b^2 \int_{-\infty}^{+\infty} dt' \Delta\alpha(t')\psi(t')\delta(t-t')$$

$$= -\omega_b^2 \int_{-\infty}^{+\infty} dt' \Delta\alpha(t')\psi(t')\left(\frac{d^2}{dt^2} + \omega_b^2\right)G(t,t'), \tag{S3}$$

which gives:

$$\psi_{tot}(t) = \psi_{inc}(t) - \omega_b^2 \int_{-\infty}^{+\infty} dt' \Delta\alpha(t')\psi_{tot}(t')G(t,t')$$

$$= \psi_{inc}(t) - \omega_b^2 \int_{-\infty}^{+\infty} dt' \Delta\alpha(t')G(t,t')\left\{\psi_{inc}(t') - \omega_b^2 \int_{-\infty}^{+\infty} dt'' \Delta\alpha(t'')G(t',t'')[\ldots]\right\}. \tag{S4}$$

Up to the first-order Born approximation, the scattered field is expressed as:

$$\psi_{sca}(t) \simeq -\omega_b^2 \int_{-\infty}^{+\infty} dt' \Delta\alpha(t')G(t,t')\psi_{inc}(t'). \tag{S5}$$

## Note S2. Causal Green's function

In source-free electromagnetic systems with time-varying $\varepsilon(t) = \alpha^{-1}(t)$, the temporal boundary conditions are derived from Maxwell's equations:

$$\frac{1}{\mu_0}\nabla \times \mathbf{B} = \frac{\partial \mathbf{D}}{\partial t}, \qquad \frac{\alpha(t)}{\varepsilon_0}\nabla \times \mathbf{D} = -\frac{\partial \mathbf{B}}{\partial t}. \tag{S6}$$

By integrating both sides of the above equations in the infinitesimal vicinity of any temporal boundary at $t = t'$:

$$\mathbf{D}(\mathbf{r},t'+0) - \mathbf{D}(\mathbf{r},t'-0) = \lim_{\Delta t \to 0}\int_{t'-\Delta t}^{t'+\Delta t} dt \frac{\partial \mathbf{D}}{\partial t} = \frac{1}{\mu_0}\lim_{\Delta t \to 0}\int_{t'-\Delta t}^{t'+\Delta t} dt \nabla \times \mathbf{B} = 0,$$

$$\mathbf{B}(\mathbf{r},t'+0) - \mathbf{B}(\mathbf{r},t'-0) = \lim_{\Delta t \to 0}\int_{t'-\Delta t}^{t'+\Delta t} dt \frac{\partial \mathbf{B}}{\partial t} = -\frac{1}{\varepsilon_0}\lim_{\Delta t \to 0}\int_{t'-\Delta t}^{t'+\Delta t} dt \alpha(t)\nabla \times \mathbf{D} = 0, \tag{S7}$$

the following two fields are derived:

$$\mathbf{D}(\mathbf{r},t) = \hat{\mathbf{x}}\psi(t)\exp(ikz)$$

$$\mathbf{B}(\mathbf{r},t) = \hat{\mathbf{y}}\frac{i\mu_0}{k}\frac{d\psi(t)}{dt}\exp(ikz) \tag{S8}$$

which should be continuous at any discontinuities of $\alpha(t)$, unless there are $\delta$-like spikes with finite areas.

According to the causality of temporal systems, the analytic expression of the Green's function can be derived from the following ansatz:

$$G(t,t') = \begin{cases} 0 & (t < t') \\ c_1 e^{-i\omega_b(t-t')} + c_2 e^{+i\omega_b(t-t')} & (t \geq t') \end{cases}. \tag{S9}$$

Incorporating the above boundary conditions: (1) the continuity of $G(t, t')$ (i.e., $\mathbf{D}$ field) and (2) the discontinuous jump of $\partial_t G(t, t')$ (i.e., $\mathbf{B}$ field) with the amount of the area of the delta function at $t = t'$:

$$G(t,t') = 0 = c_1 + c_2$$

$$\left.\frac{\partial}{\partial t}G(t,t')\right|_{t=t'+0} - \left.\frac{\partial}{\partial t}G(t,t')\right|_{t=t'-0} = 1 = -i\omega_b(c_1 - c_2) - 0. \tag{S10}$$

Therefore, $c_1 = -c_2 = i/2\omega_b$, and Eq. (3) in the main text is obtained.

On the other hand, the Green's function of the Helmholtz equation [Eq. (S2)] can be obtained alternatively using a Fourier transform by supposing that:

$$G(t,t') = \int_{-\infty}^{\infty}\frac{d\omega}{2\pi}\tilde{G}(\omega,t')e^{-i\omega t} \quad \text{and} \quad \tilde{G}(\omega,t') = \int_{-\infty}^{\infty} dt G(t,t')e^{+i\omega t}, \tag{S11}$$

Eq. (S2) is expressed in the frequency domain as:

$$(\omega_b^2 - \omega^2)\tilde{G}(\omega,t') = e^{i\omega t'}, \qquad \therefore \tilde{G}(\omega,t') = -\frac{e^{i\omega t'}}{\omega^2 - \omega_b^2}. \tag{S12}$$

The Green's function in the time domain is then obtained again by applying the inverse Fourier transformation to the above equation, while the result of the integration in the inverse Fourier transformation can be different according to the physical systems of interest, as shown in the retarded, advanced, or Feynman propagators for the Klein-Gordon equation[1]. In this respect, among the possible approximations for locating the poles ($\omega = \pm \omega_b$) in Eq. (S12) in relation to Cauchy's residue theorem (above or below the real axis, Fig. S1), we choose both poles below the real axis to guarantee the causality condition, $G(t < t', t') = 0$:

$$G(t,t') = -\int_{-\infty}^{\infty} \frac{d\omega}{2\pi} \frac{e^{-i\omega(t-t')}}{\omega^2 - \omega_b^2}$$

$$= -\lim_{\varepsilon \to 0+} \int_{-\infty}^{\infty} \frac{d\omega}{2\pi} \frac{e^{-i\omega(t-t')}}{(\omega - \omega_b + i\varepsilon)(\omega + \omega_b + i\varepsilon)}$$

$$= -\Theta(t-t') \lim_{\varepsilon \to 0+} \frac{2\pi i}{2\pi} \left[ -\operatorname*{Res}_{\omega=\omega_b - i\varepsilon} \frac{e^{-i\omega(t-t')}}{(\omega-\omega_b+i\varepsilon)(\omega+\omega_b+i\varepsilon)} - \operatorname*{Res}_{\omega=-\omega_b-i\varepsilon} \frac{e^{-i\omega(t-t')}}{(\omega-\omega_b+i\varepsilon)(\omega+\omega_b+i\varepsilon)} \right] \quad \text{(S13)}$$

$$= \frac{1}{\omega_b} \Theta(t-t') \left[ -\frac{e^{-i\omega_b(t-t')}}{2i} + \frac{e^{i\omega_b(t-t')}}{2i} \right] = \frac{1}{\omega_b} \sin\left[\omega_b(t-t')\right] \Theta(t-t').$$

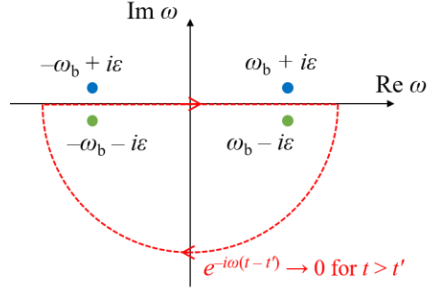

**Fig. S1. Schematic of the contour integration in Eq. (S13)**. clockwise half-infinite contour for $t > t'$ (red dashed line) and selected poles (green dots) below the real axis, compared to the other candidates (blue dots) above the real axis.

## Note S3. Numerical validation of the Born approximation

The Born approximation is valid when the first-order amplitude of the scattering wave is much smaller than the incident wave amplitude[2], which also results in the convergence of the Born series. Therefore, it is necessary to obtain:

$$\left| \omega_b^2 \int_{-\infty}^{+\infty} dt' \Delta\alpha(t') G(t,t') \psi_{\text{inc}}(t') \right| = \left| \omega_b \int_{-\infty}^{t} dt' \Delta\alpha(t') \sin[\omega_b(t-t')] e^{-\omega_b t'} \right| \ll 1. \quad (S14)$$

We introduce two parameters regarding the perturbation $\Delta\alpha(t)$: the fluctuation scale $\delta$ and the correlation time $\sigma$, which determine the upper bound of the fluctuation $|\Delta\alpha(t)| \leq \delta$ and the lower bound of a nonzero spectral component, as $\Delta\alpha(\omega > \sigma^{-1}) = 0$, respectively. We also assume a finite-time modulation with time $T$, such that $\Delta\alpha(t) = 0$ for $t < 0$ or $t > T$. Sufficient conditions for Eq. (S14) are derived as follows:

$$\left| \omega_b \int_{-\infty}^{t} dt' \Delta\alpha(t') \sin[\omega_b(t-t')] e^{-\omega_b t'} \right| \leq \omega_b \int_{-\infty}^{t} dt' |\Delta\alpha(t')| \leq \omega_b T \delta \ll 1, \quad (S15)$$

and

$$\sigma^{-1} \ll 2\omega_b, \quad (S16)$$

each describing the weak and slowly varying perturbation.

The above analysis can be verified using numerical assessments. Supposing $T = 10 t_0$ and the above $\delta$ and $\sigma$ as the square root of amplitude and the standard deviation of the Gaussian correlation function: $S(\Delta t) = \delta^2 \exp[-(\Delta t/\sigma)^2/2]$, we estimate the error of the Born approximation compared to the numerically calculated ground-truth results (transfer matrix method, TMM) for various combinations of $(\sigma, \delta)$. For this, we quantify the error with the mean absolute percentage error (MAPE) over the time period ($0 < t < T$):

$$\text{MAPE} = \frac{1}{T} \int_0^T dt \frac{\left| \psi_{\text{sca}}^{(\text{Born})}(t) \right|^2 - \left| \psi_{\text{sca}}^{(\text{TMM})}(t) \right|^2}{\left| \psi_{\text{sca}}^{(\text{TMM})}(t) \right|^2}. \quad (S17)$$

Figure S2 shows the ensemble-averaged MAPE for $10^3$ realizations per ensemble, demonstrating the acceptable regime of the two parameters, as marked by the red dashed line in Fig. S2c. For example, each of the two points (a, b) in Fig. S2c exemplifies the good and bad approximation results, respectively. While the approximated (blue dashed line) and ground-truth (purple solid line) scattering intensities are well matched in Fig. S2a, there is a noticeable error between the two results in Fig. S2b.

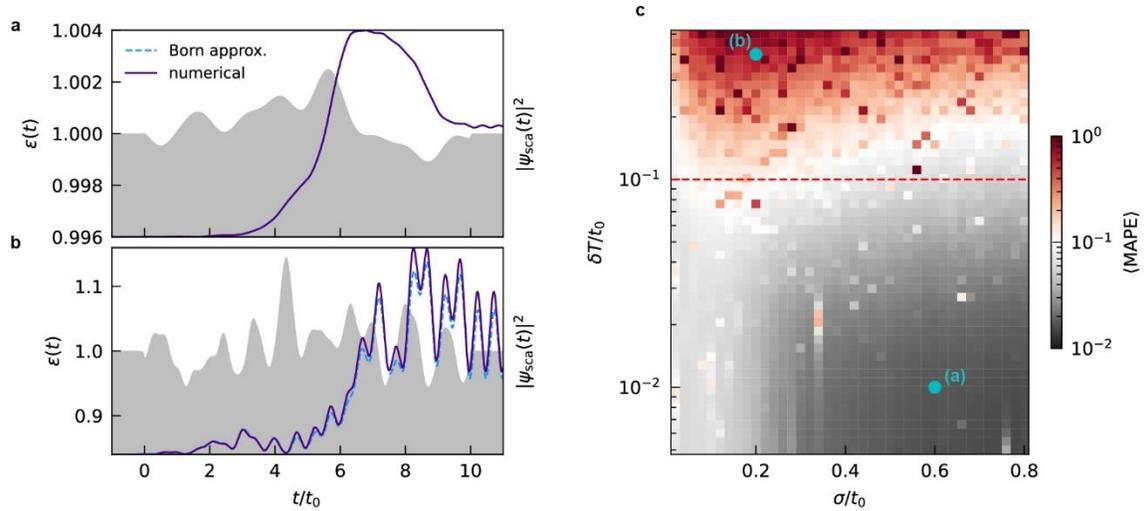

**Fig. S2. Numerical validations of the Born approximation. a, b,** Two realizations for $(\sigma/t_0, \delta T/t_0) = (0.6, 0.01)$ **(a)** and $(0.2, 0.4)$ **(b)** with $T = 10 t_0$. **c,** Ensemble-averaged MAPE for various combinations of $(\sigma/t_0, \delta T/t_0)$. $10^3$ realizations per ensemble.

**Note S4. Directional separation of scattering waves**

Combining Eqs (2) and (3) in the main text with the incident wave $\psi_{\text{inc}}(t) = \exp(-i\omega_b t)$, the separated form of the scattering wave in Eq. (4) is derived as follows:

$$\begin{aligned}
\psi_{\text{sca}}(t) &= -\omega_b^2 \int_{-\infty}^{\infty} dt' \Delta\alpha(t') G(t,t') \psi_{\text{inc}}(t') \\
&= -\omega_b \int_{-\infty}^{t} dt' \Delta\alpha(t') \sin[\omega_b(t-t')] e^{-i\omega_b t'} \\
&= -\frac{\omega_b}{2i} \int_{-\infty}^{t} dt' \Delta\alpha(t') \left[ e^{i\omega_b(t-t')} - e^{-i\omega_b(t-t')} \right] e^{-i\omega_b t'} \\
&= \frac{\omega_b}{2i} \int_{-\infty}^{t} dt' \Delta\alpha(t') \left[ e^{-i\omega_b t} - e^{i\omega_b(t-2t')} \right] = \frac{\omega_b}{2i} \left[ e^{-i\omega_b t} \int_{-\infty}^{t} dt' \Delta\alpha(t') - e^{+i\omega_b t} \int_{-\infty}^{t} dt' e^{-i\omega_b 2t'} \Delta\alpha(t') \right],
\end{aligned} \tag{S18}$$

where the resulting two terms in Eq. (S18) are the forward and backward scattering waves, respectively, as follows:

$$\begin{aligned}
\psi_{\text{sca,FW}}(t) &= \frac{\omega_b}{2i} e^{-i\omega_b t} \int_{-\infty}^{t} dt' \Delta\alpha(t') \\
\psi_{\text{sca,BW}}(t) &= -\frac{\omega_b}{2i} e^{+i\omega_b t} \int_{-\infty}^{t} dt' e^{-i\omega_b 2t'} \Delta\alpha(t').
\end{aligned} \tag{S19}$$

## Note S5. Ensemble-averaged forward and backward powers

The scattered power is obtained by the square of the scattering wave $\psi_{sca}(t) = \psi_{sca,FW}(t) + \psi_{sca,BW}(t)$ in Eq. (S19):

$$|\psi_{sca}(t)|^2 = |\psi_{sca,FW}(t) + \psi_{sca,BW}(t)|^2 = |\psi_{sca,FW}(t)|^2 + |\psi_{sca,BW}(t)|^2 + 2\text{Re}\left[\psi^*_{sca,FW}(t)\psi_{sca,BW}(t)\right], \quad (S20)$$

where $P_{FW} = |\psi_{sca,FW}(t)|^2$ and $P_{BW} = |\psi_{sca,BW}(t)|^2$ are the scattering powers in the forward and backward directions, respectively, and the last term $2\text{Re}[\psi^*_{sca,FW}(t)\psi_{sca,BW}(t)]$ corresponds to the interference between the two waves. The scattering power in the forward direction is calculated from Eq. (S19) as follows:

$$\begin{aligned}
P_{FW}(t) &= |\psi_{sca,FW}(t)|^2 \\
&= \left|\frac{\omega_b}{2i} e^{-i\omega_b t} \int_{-\infty}^{t} dt' \Delta\alpha(t')\right|^2 = \frac{\omega_b^2}{4}\left[\int_{-\infty}^{t} dt' \Delta\alpha(t')\right]^*\left[\int_{-\infty}^{t} dt' \Delta\alpha(t')\right] \\
&= \frac{\omega_b^2}{4}\int_{-\infty}^{t} dt_1' \int_{-\infty}^{t} dt_2' \Delta\alpha^*(t_1')\Delta\alpha(t_2'),
\end{aligned} \quad (S21)$$

where the primed variables $t_{1,2}'$ denote the coordinates of two independent scatterers. The ensemble-averaged forward scattering power is then expressed using the two-point correlation function $\hat{S}(t_1', t_2') \equiv \langle\Delta\alpha^*(t_1')\Delta\alpha(t_2')\rangle$:

$$\langle P_{FW}(t)\rangle = \frac{\omega_b^2}{4}\int_{-\infty}^{t} dt_1' \int_{-\infty}^{t} dt_2' \langle\Delta\alpha^*(t_1')\Delta\alpha(t_2')\rangle = \frac{\omega_b^2}{4}\int_{-\infty}^{t} dt_1' \int_{-\infty}^{t} dt_2' \hat{S}(t_1', t_2'). \quad (S22)$$

Setting the finite range of perturbation (*i.e.*, $\hat{S}(t_1', t_2') = 0$ for $t_{1,2}' < 0$ or $t_{1,2}' > T$), the forward scattering power after the perturbation ($t > T$) becomes:

$$\langle P_{FW}(t > T)\rangle = \frac{\omega_b^2}{4}\int_0^T dt_1' \int_0^T dt_2' \hat{S}(t_1', t_2') = \frac{\omega_b^2}{4}\int_0^T dt_1' \int_0^T dt_2' S(t_1' - t_2') \quad (S23)$$

with statistical homogeneity along the temporal axis: $\hat{S}(t_1', t_2') = S(t_1' - t_2')$. Similarly, the backward scattering power is expressed as

$$\langle P_{BW}(t > T)\rangle = \frac{\omega_b^2}{4}\int_0^T dt_1' \int_0^T dt_2' S(t_1' - t_2')e^{2i\omega_b(t_1' - t_2')}. \quad (S24)$$

### Note S6. Long-range approximation

When the typical correlation time is much less than the finite duration $T$, the two-point correlation function $\hat{S}(t_1', t_2')$ can be expressed as a sparse matrix with vanishing off-diagonal components. In this sufficiently long perturbation (large $T$), the exact results from the double integral in Eqs. (S23) and (S24) can be approximated as follows (Fig. S3):

$$\langle P_{\text{FW}}(T)\rangle = \frac{\omega_b^2}{4}\int_0^T dt_1'\int_0^T dt_2' S(t_1'-t_2')$$

$$\simeq \frac{\omega_b^2}{4}\int_0^T dt_1'\int_{-\infty}^{\infty} dt_2' S(t_1'-t_2') = \frac{\omega_b^2}{4}\int_0^T dt_1'\int_{-\infty}^{\infty} d\Delta t\, S(\Delta t)$$

$$= \frac{\omega_b^2 T}{4}\int_{-\infty}^{\infty} d\Delta t\, S(\Delta t) = \frac{\omega_b^2 T}{4} S(\omega=0),$$

$$\langle P_{\text{BW}}(T)\rangle = \frac{\omega_b^2}{4}\int_0^T dt_1'\int_0^T dt_2' S(t_1'-t_2')e^{2i\omega_b(t_1'-t_2')}$$

$$\simeq \frac{\omega_b^2}{4}\int_0^T dt_1'\int_{-\infty}^{\infty} dt_2' S(t_1'-t_2')e^{2i\omega_b(t_1'-t_2')} = \frac{\omega_b^2}{4}\int_0^T dt_1'\int_{-\infty}^{\infty} d\Delta t\, S(\Delta t)e^{2i\omega_b \Delta t}$$

$$= \frac{\omega_b^2 T}{4}\int_{-\infty}^{\infty} d\Delta t\, S(\Delta t)e^{2i\omega_b \Delta t} = \frac{\omega_b^2 T}{4} S(\omega=2\omega_b).$$

(S25)

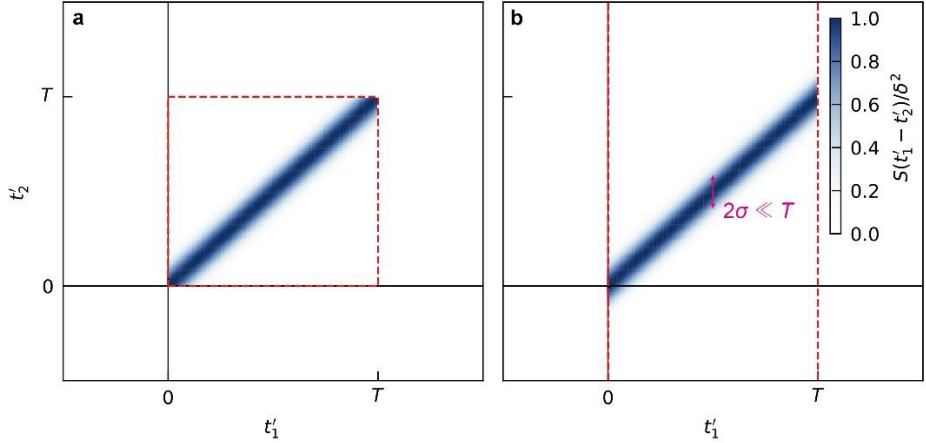

**Fig. S3. Condition of long-range approximation. a**, Illustration of the exact 2D integration [Eq. S(23)] in the finite domain of a square shape with red dashed lines. **b**, Illustration of the approximated integration [Eq. S(25)] with the infinitely extended domain between two red dashed lines for a sufficiently large $T$ compared to the correlation time scale $\sigma$.

## Note S7. Gaussian random generation and the conditions for structure factors

Throughout our study, the random realization of the correlated disorder is obtained by the multivariate Gaussian process. That is, a correlated disorder $\mathbf{x} = L\mathbf{z}$ is expressed as a linear transform of an uncorrelated standard normal disorder $\mathbf{z}$ with an operator $L$, where $L$, $\mathbf{x}$, and $\mathbf{z}$ are defined in a finite-dimensional vector space based on the discretization of the time domain. It is observed that $L$ is the lower-triangular operator of the Cholesky decomposition [3] of covariance matrix $\Gamma$:

$$\Gamma \equiv \langle \mathbf{x}\mathbf{x}^\dagger \rangle = \langle (L\mathbf{z})(L\mathbf{z})^\dagger \rangle = \langle L\mathbf{z}\mathbf{z}^\dagger L^\dagger \rangle = L\langle \mathbf{z}\mathbf{z}^\dagger \rangle L^\dagger = LL^\dagger, \tag{S26}$$

where the component of $\Gamma$ is expressed using the correlation function $S(\Delta t)$:

$$\Gamma_{i,j} = \langle x_i x_j^* \rangle = \langle \Delta\alpha^*(t_j)\Delta\alpha(t_i) \rangle = S(t_j - t_i). \tag{S27}$$

To generate a discretized random realization $\mathbf{x} = [\Delta\alpha(t_1), \Delta\alpha(t_2), \ldots]^T$ with a given correlation function $S(\Delta t)$, it is necessary to decompose the covariance matrix $\Gamma$ with Cholesky factorization, which is derived from $S(\Delta t)$. The lower triangular matrix $L$ can then be applied to an uncorrelated standard normal vector $\mathbf{z}$.

There are a few conditions to be considered for this numerical generation of correlated disorder. First, the Hermiticity of $S(\Delta t)$ is trivial by definition:

$$\begin{aligned}S(-\Delta t) &= \langle \Delta\alpha^*(t - \Delta t)\Delta\alpha(t) \rangle = \langle \Delta\alpha^*(t)\Delta\alpha(t + \Delta t) \rangle \\ &= \langle \Delta\alpha(t)\Delta\alpha^*(t + \Delta t) \rangle^* = S^*(\Delta t),\end{aligned} \tag{S28}$$

which leads to the real-valued structure factor from:

$$\begin{aligned}S^*(\omega) &= \left[\int_{-\infty}^{\infty} d\Delta t\, S(\Delta t) e^{i\omega\Delta t}\right]^* = \int_{-\infty}^{\infty} d\Delta t\, S^*(\Delta t) e^{-i\omega\Delta t} = \int_{-\infty}^{\infty} d\Delta t\, S(-\Delta t) e^{-i\omega\Delta t} \\ &= \int_{\infty}^{-\infty} (-d\Delta t) S(\Delta t) e^{i\omega\Delta t} = \int_{-\infty}^{\infty} d\Delta t\, S(\Delta t) e^{i\omega\Delta t} = S(\omega).\end{aligned} \tag{S29}$$

The second condition regards the maximum value of the correlation function: $-S(0) \leq \mathrm{Re}\, S(\Delta t) \leq S(0)$, proven by

$$\begin{aligned}\langle |\Delta\alpha(t + \Delta t) \pm \Delta\alpha(t)|^2 \rangle &= \langle |\Delta\alpha(t + \Delta t)|^2 + |\Delta\alpha(t)|^2 \pm 2\mathrm{Re}\,\Delta\alpha^*(t + \Delta t)\Delta\alpha(t) \rangle \\ &= \langle |\Delta\alpha(t + \Delta t)|^2 \rangle + \langle |\Delta\alpha(t)|^2 \rangle \pm 2\mathrm{Re}\langle \Delta\alpha^*(t + \Delta t)\Delta\alpha(t) \rangle \\ &= 2S(0) \pm 2\mathrm{Re}\, S(\Delta t) \geq 0.\end{aligned} \tag{S30}$$

Finally, we note that semipositive $S(\omega)$ is also required as a sufficient condition to generate a covariance matrix. Suppose that $0 \leq S(\omega)$, the positive semidefiniteness of the covariance matrix $\Gamma$ is derived from the following relationship:

$$\begin{aligned}\mathbf{y}^\dagger \Gamma \mathbf{y} &= \sum_{i,j} y_i^* \Gamma_{i,j} y_j = \sum_{i,j} y_i^* y_j S(t_j - t_i) \\ &= \sum_{i,j} y_i^* y_j \int_{-\infty}^{\infty} \frac{d\omega}{2\pi} S(\omega) e^{-i\omega(t_j - t_i)} \\ &= \int_{-\infty}^{\infty} \frac{d\omega}{2\pi} S(\omega) \sum_{i,j} y_i^* y_j e^{-i\omega(t_j - t_i)} \\ &= \int_{-\infty}^{\infty} \frac{d\omega}{2\pi} S(\omega) \left|\sum_j y_j e^{-i\omega t_j}\right|^2 \geq 0, \quad \forall \mathbf{y},\end{aligned} \tag{S31}$$

where $\mathbf{y} = [y_1, y_2, \ldots]^T$ is an arbitrary vector. Every Hermitian and positive-definite matrix has a unique Cholesky decomposition[3], which consequently leads to the successful generation of a realization from a given positive structure factor.

## Note S8. Design of structure factors for target forward and backward scatterings

While the two values of a structure factor $S(\omega = 0)$ and $S(\omega = 2\omega_0)$ independently determine the statistical scattering amplitudes in the forward and backward directions, respectively, the overall shape of the structure factor beyond the two function values, including area, continuity, and smoothness, should be considered simultaneously for the generation of realizations. For example, the area of structure factors:

$$2\pi S(\Delta t = 0) = \int_{-\infty}^{\infty} d\omega S(\omega) \tag{S32}$$

should be controlled for removing the scaling effect of disorder fluctuation in realizations. In addition, the continuous and smooth shape of a target structure factor is also necessary for stable numerical generation of realizations.

Considering the above points, for example, the structure factor for forward control $S_{FW}(\omega)$ can be designed with the following considerations. First, the structure factor is factorized with $[1 - (\omega/2\omega_0)^2]$ to satisfy $S_{FW}(\pm 2\omega_0) = 0$, resulting in the expression:

$$S_{FW}(\omega) = \left[1 - \left(\frac{\omega}{2\omega_0}\right)^2\right]^n [S_0 + X(\omega)], \tag{S33}$$

where $X(\omega)$ is an auxiliary function such that $X(\omega = 0) = 0$ to satisfy $S_{FW}(\omega = 0) = S_0$. Note that the order $n$ should not be less than 2, which is necessary for smoothening the shape of the structure factor in the vicinity of $\omega = 2\omega_0$, with $S_{FW}'(\pm 2\omega_0) = 0$ ($n = 2$ in the main text). In addition, $X(\omega)$ should also be factorized by $\omega^m$, and for smoothening, we choose $m = 2$. Finally, in the setting $X(\omega) = K_{FW}(4\omega/\omega_0)^2$ for a dependent variable $K_{FW}(S_0)$, we apply the integral condition:

$$\int_{-\infty}^{\infty} \frac{d\omega}{2\pi} S(\omega) = S(\Delta t = 0) = \delta^2, \tag{S34}$$

which results in:

$$S_{FW}(\omega; S_0) = \left[1 - \left(\frac{\omega}{2\omega_0}\right)^2\right]^2 \left[S_0 + K_{FW}(S_0)\left(\frac{4\omega}{\omega_0}\right)^2\right],$$

$$K_{FW}(S_0) = \frac{1}{4}\left(\frac{105\pi\delta^2}{256\omega_0} - \frac{7S_0}{16}\right). \tag{S35}$$

Similarly, we design a structure factor $S_{BW}(\omega)$ for the backward scattering as:

$$S_{BW}(\omega; S_{2\omega}) = \left(\frac{\omega}{2\omega_0}\right)^2 \left(3 - \frac{|\omega|}{\omega_0}\right)\left[S_{2\omega} + 4K_{BW}(S_{2\omega})\left(\frac{|\omega|}{\omega_0} - 2\right)^2\right], \tag{S36}$$

where:

$$K_{BW}(S_{2\omega}) = \frac{10\pi\delta^2}{27\omega_0} - \frac{5S_{2\omega}}{8}. \tag{S37}$$

## Note S9. Estimation of $S(\omega)$ for generated realizations

To verify whether the generated realizations describe the target structure factor, we calculate the estimation of the structure factor based on the ergodicity in the correlation functions:

$$S^{(\text{est})}(0 \leq \Delta t \leq T; N_{\text{ens}}) = \frac{1}{N_{\text{ens}}} \sum_{i=1}^{N_{\text{ens}}} \frac{1}{T - \Delta t} \int_0^{T-\Delta t} dt' \Delta \alpha_i^*(t' + \Delta t) \Delta \alpha_i(t'),$$

$$S^{(\text{est})}(\omega; N_{\text{ens}}) = \int_0^T d(\Delta t) S^{(\text{est})}(\Delta t)\left(e^{i\omega \Delta t} + e^{-i\omega \Delta t}\right),$$

(S38)

where $S^{(\text{est})}(\Delta t; N_{\text{ens}})$ is the estimation of the ensemble-averaged correlation function for total $N_{\text{ens}}$ different realizations of $\Delta\alpha_i(t)$ ($i = 1, 2, \ldots, N_{\text{ens}}$), and $S^{(\text{est})}(\omega; N_{\text{ens}})$ is the Fourier transform for the symmetric correlation function $S(-\Delta t) = S(\Delta t)$. In Fig. S4, several examples of target structure factors in Eqs. (S35) and (S36) and their statistical estimations are displayed. As $N_{\text{ens}}$ increases, the estimated structure factors approach the corresponding target structure factors. Notably, the order of $N_{\text{ens}} = 10^4$ leads to almost precise coincidence between the target and estimated structure factors. Therefore, we choose $N_{\text{ens}} = 10^4$ as the realization number per ensemble for statistical analysis in our study.

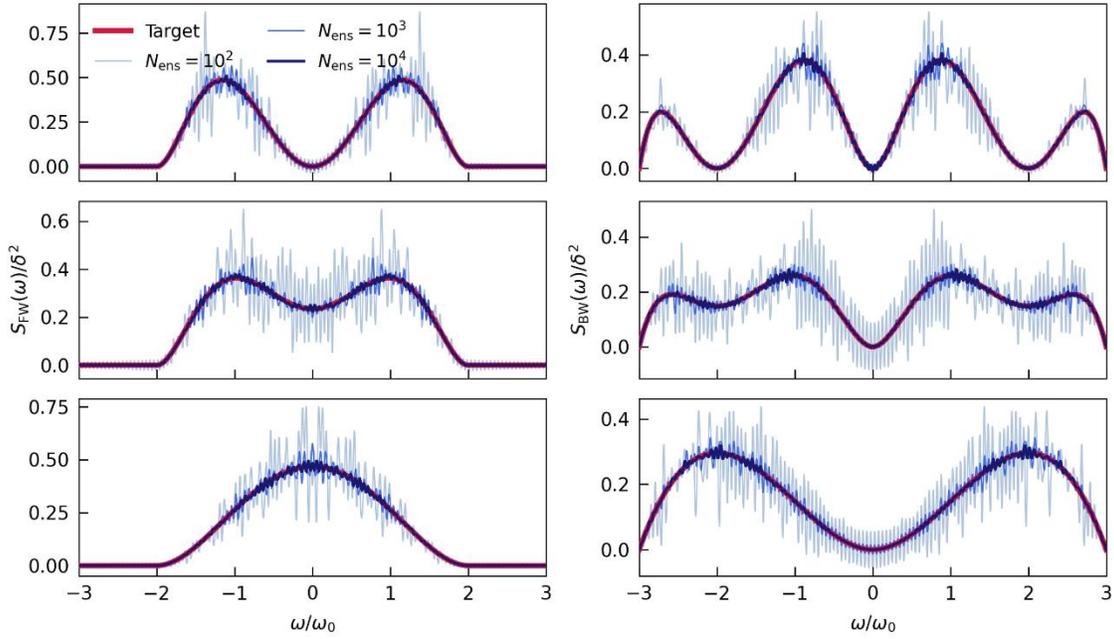

**Fig. S4. Estimation of the structure factors.** (left) $S_{\text{FW}}^{(\text{est})}(\omega)$ and (right) $S_{\text{BW}}^{(\text{est})}(\omega)$ compared to the corresponding target structure factors $S_{\text{FW,BW}}(\omega)$ for different values of design parameters, $S_{0,2\omega} = 0$ (top), max/2 (middle), and max (bottom).

**Note S10. Details of $S_C(\omega)$ and $S_P(\omega)$**

We first introduce the function $B(x)$:

$$B(x) = \begin{cases} (x^2 - 1)^2, & |x| \leq 1 \\ 0, & \text{otherwise} \end{cases}, \quad (S39)$$

which is continuous and of the $C^1$ class, thereby continuously differentiable. We utilize this function to describe both sharp Bragg peaks and near-Poisson broad bumps with different bandwidths.

First, the structure factor $S_P(\omega)$ for the near-Poisson structure factor is expressed with a modified width,

$$S_P(\omega) \sim B\left(\frac{\omega}{\Delta\omega}\right), \quad (S40)$$

where $\Delta\omega = 2\omega_0$ is the standard bandwidth of the broadband bump.

Next, the structure factor $S_C(\omega)$ for the crystal is expressed as:

$$S_C(\omega) \sim \text{sech}\left(\frac{\omega}{\Delta\omega/4}\right) \sum_{\omega_c} B\left(\frac{\omega - \omega_c}{\Delta\omega/40}\right), \quad (S41)$$

where $\omega_c$ is the summation index over [$\pm 0.5\omega_0, \pm\omega_0, \pm 1.5\omega_0, \pm 2.5\omega_0, \pm 3\omega_0, \pm 3.5\omega_0, \pm 4\omega_0$] for the centre frequencies of Bragg peaks with reduced bandwidth ($\Delta\omega/4$) compared to $S_P(\omega)$, and the sech function represents the envelope function for different heights of the Bragg peaks. We note that both $S_P(\omega)$ and $S_C(\omega)$ are numerically normalized so that:

$$\int_{-4\omega_0}^{4\omega_0} \frac{d\omega}{2\pi} S_{C,P}(\omega) = S_{C,P}(\Delta t = 0) = \delta^2. \quad (S42)$$

In this setting, the structure factor of Eq. (7) in the main text can possess the same onsite correlation equal to $\delta^2$ regardless of $\xi$.